\documentclass[aps,nofootinbib,superscriptaddress,twocolumn,prd,10pt,floatfix]{revtex4-1}

\usepackage{graphicx}
\usepackage{amsmath,amssymb}
\usepackage{hyperref}
\usepackage{braket}
\usepackage{booktabs}
\usepackage{subfigure}
\usepackage[dvipsnames]{xcolor}

\newcommand{\fDR}{f_{\rm DR}}
\newcommand{\fACDM}{f_{\rm AcDM}}

\hypersetup{
    colorlinks=true,
    linkcolor=red,
    citecolor=blue,
}

\newcommand{\nl}{\\ \indent}

\allowdisplaybreaks
\begin{document}

\title{Partially Acoustic Dark Matter Cosmology and Cosmological Constraints}

\author{Marco Raveri}
\affiliation{Kavli Institute for Cosmological Physics, Department of Astronomy \& Astrophysics, Enrico Fermi Institute, The University of Chicago, Chicago, IL 60637, USA}
\affiliation{Institute Lorentz, Leiden University, PO Box 9506, Leiden 2300 RA, The Netherlands}
\author{Wayne Hu}
\affiliation{Kavli Institute for Cosmological Physics, Department of Astronomy \& Astrophysics, Enrico Fermi Institute, The University of Chicago, Chicago, IL 60637, USA}
\author{Timothy Hoffman}
\affiliation{Kavli Institute for Cosmological Physics, Department of Physics, Enrico Fermi Institute, The University of Chicago, Chicago, IL 60637, USA}
\author{Lian-Tao Wang}
\affiliation{Kavli Institute for Cosmological Physics, Department of Physics, Enrico Fermi Institute, The University of Chicago, Chicago, IL 60637, USA}

\begin{abstract}
Observations of the cosmic microwave background (CMB) together with weak lensing measurements of the clustering of large scale cosmological structures and local measurements of the Hubble constant pose a challenge to the standard $\Lambda$CDM cosmological model. 
On one side CMB observations imply a Hubble constant that is lower than local measurements and an amplitude of the lensing signal that is higher than direct measurements from weak lensing surveys. 
We investigate a way of  relieving these tensions by adding dark radiation tightly coupled to an acoustic part of the dark matter sector and compare it to massive neutrino solutions. While these models offer a way of separately relieving the Hubble and weak lensing tensions they are prevented from fully accommodating both at the same time since the CMB requires additional cold dark matter when adding acoustic dark matter or massive neutrinos to preserve the same sharpness  of the
acoustic peaks which counteracts the desired growth suppression.
\end{abstract}

\maketitle

\section{Introduction}
The description of the Universe based on a cosmological constant ($\Lambda$) and cold dark matter (CDM) has provided an excellent model to fit cosmological observations since the discovery of cosmic acceleration~\cite{Riess:1998cb,Perlmutter:1998np}.
However, as experimental sensitivity increases, tensions of various degrees of significance arise between different observations.
Measurements of the cosmic microwave background (CMB) from the {\it Planck} satellite~\cite{Adam:2015rua} imply a Hubble constant in $\Lambda$CDM 
that is significantly lower than direct measurement from the local Hubble flow~\cite{Riess:2016jrr} and strong lensing time delays~\cite{Bonvin:2016crt}. The former results in an estimate of $H_0 = 67.51 \pm 0.64 \, {\rm km}\,{\rm s}^{-1}\,{\rm Mpc}^{-1}$~\cite{Ade:2015xua} which is $3.1 \sigma$ and $2.1 \sigma$ lower than the latter two that give $H_0 = 73.24 \pm 1.74 \, {\rm km}\,{\rm s}^{-1}\,{\rm Mpc}^{-1}$ and $H_0 = 72.8 \pm 2.4 \, {\rm km}\,{\rm s}^{-1}\,{\rm Mpc}^{-1}$.
On the other hand measurements of weak gravitational lensing (WL), in particular from the Canada-France-Hawaii Telescope Lensing Survey (CFHTLenS)~\cite{Heymans:2013fya} and Kilo Degree Survey (KiDS)~\cite{Hildebrandt:2016iqg}, result in an estimate of the amplitude of the lensing signal that is low compared to the one derived from {\it Planck} $\Lambda$CDM cosmology. 
While it is possible that these tensions point to systematic errors in these data sets, none have been uncovered to date. \nl
This  situation motivated the development of models aimed at solving these tensions by either modifying the neutrino sector \cite{Wyman:2013lza,Adhikari:2016bei,Canac:2016smv,Feng:2017nss}, the  gravitational sector~\cite{Hamann:2013iba,Costanzi:2014tna,Ade:2015rim,Hu:2015rva,Joudaki:2016kym,Peirone:2017lgi,Zhao:2017cud}, or the  dark matter sector~\cite{Buen-Abad:2015ova, Enqvist:2015ara, Lesgourgues:2015wza, Poulin:2016nat, Ko:2016uft, Ko:2016fcd}.
We shall here focus on one attempt, belonging to the latter groups, that was proposed in~\cite{Chacko:2016kgg} and dubbed Partially Acoustic  Dark Matter (PAcDM).
\begin{figure*}[th]
\centering
\includegraphics{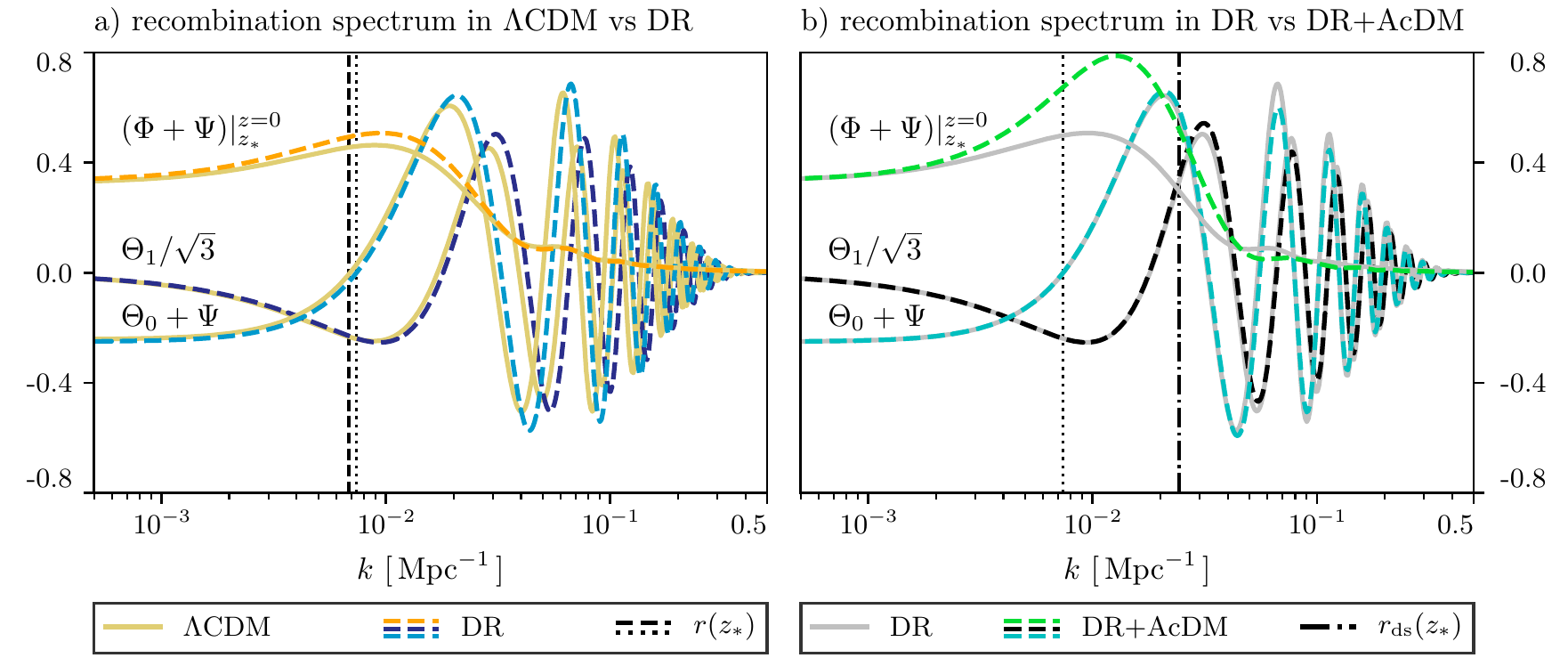}
\caption{
The recombination spectrum in $k$-space in units of amplitude of primordial comoving curvature perturbation. Different lines correspond to different physical effects and models, as shown in figure and legend. The vertical dashed and dotted lines show the comoving horizon at recombination ($z_*$) for the $\Lambda$CDM and DR models. The dot-dashed line in Panel (b) shows the comoving dark sound horizon at recombination.
The $\Lambda$CDM model is defined by $\Omega_b=0.05$, $\Omega_c=0.26$, $h=0.67$, the DR model has $f_{\rm DR}=0.6$ and the DR+AcDM model further adds $f_{\rm AcDM}=0.6$ while leaving unchanged all other parameters. 
}
\label{Fig:RecombinationSpectrum}
\end{figure*}
In addition to the standard components of the $\Lambda$CDM model PAcDM considers a Dark Radiation (DR) species, that significantly contributes to the energy density of the universe around the time of matter-radiation equality. This component is tightly self-coupled at all times so that it behaves like a fluid.
In addition it includes a subdominant dark matter component that is strongly interacting with the DR fluid.
Such dark matter would undergo acoustic oscillations below the effective dark sound horizon and is therefore referred to as Acoustic  Dark Matter (AcDM).   \nl
We show that the model has two viable corners of parameter space: one that is helping in relieving the tension between CMB and WL measurements, while not improving the discrepancy with local Hubble measurements; a second one that relieves the tension with $H_0$ determinations but has only marginal decrease in growth, corresponding to a limited efficiency at solving the WL tension. 
The combination of parameters that could solve the joint tension, as well as the benchmark parameters proposed in~\cite{Chacko:2016kgg}, are disfavored by the temperature spectrum of the CMB, as measured by the {\it Planck} satellite due mainly to conflicting requirements for cold dark matter abundance. \nl
The phenomenology of this model is similar to, but in detail different, from models involving extra energy density in massive neutrinos.
We therefore compare and contrast results for PAcDM with active massive neutrinos (M$\nu$) (see e.g.~\cite{Lesgourgues:2006nd} for a review)  and sterile neutrinos~\cite{GellMann:1980vs,PhysRevLett.44.912,Dodelson:1993je} (S$\nu$) extensions of $\Lambda$CDM. 
Overall we find that the PAcDM model does have similar performances to neutrino models, while being slightly better at reducing the WL tension alone and slightly worse for the $H_0$ tension alone. For the full data set combination that we consider, PAcDM is slightly better and reaches a draw in an evidence based comparison with the $\Lambda$CDM model with the two neutrino models being slightly disfavored. \nl
This paper is organized as follows. 
In Sec.~\ref{Sec:Cosmology} we discuss the cosmological phenomenology of the PAcDM model, with a particular focus on the effect on the CMB temperature spectrum.
In Sec.~\ref{Sec:DataSets} we outline the details of the data sets and the tools that we use to test the model against cosmological observations.
In Sec.~\ref{Sec:Constraints} we show the result of this comparison and summarize our conclusions in Sec.~\ref{Sec:Conclusions}.

\section{Partially Acoustic Dark Matter Cosmology} \label{Sec:Cosmology}

In this section we briefly review the phenomenological implications of PAcDM and discuss its effect on the temperature spectrum of the CMB and the clustering of large scale cosmological structures.
Inspired by this discussion, we shall develop a new basis for the model parameters that better isolates its observational differences with $\Lambda$CDM. This will aid the interpretation of the data constraints in the next sections.
Following~\cite{Chacko:2016kgg}, we define the PAcDM model by means of specifying the tightly coupled dark radiation (DR) and acoustic dark matter (AcDM) fractional densities in units of massless neutrino density and cold dark matter density:
\begin{align} \label{Eq:BaseParamsDef}
f_{\rm DR}    =& \frac{\rho_{\rm DR}}{\rho_{\nu}} \equiv \frac{\Delta N_{\rm eff}}{N_{\rm eff}} \,, \nonumber\\
f_{\rm AcDM} =& \frac{\rho_{\rm AcDM}}{\rho_c + \rho_{\rm AcDM}} \,.
\end{align}
Evaluated today these two quantities, in addition to those of the baseline $\Lambda$CDM will serve as the cosmological parameters for the model.
When not otherwise stated we shall use $N_{\rm eff}=3.046$ as the default value when defining DR fractional density. \nl
In all the model considered the addition of extra dark radiation has an effect on the cosmological expansion history, shifting toward later times matter radiation equality.
This change in the expansion history changes the calibration of the acoustic scale and hence the inference of the angular diameter distance to recombination and Hubble constant.
Adding AcDM does not affect the background expansion history as the total matter density remains the same as we change $f_{\rm AcDM}$. \nl
The addition of a DR fluid, coupled or not coupled to AcDM, would alter the behavior of perturbations as well, leaving a significant imprint on the CMB temperature spectrum.
We shall use Fig.~\ref{Fig:RecombinationSpectrum} to aid our interpretation of these effects.
There we show the monopole, $\Theta_0$, of the radiation field, its dipole $\Theta_1$ at the redshift of recombination $z_*$ and the difference in gravitational potentials between recombination and today $(\Phi+\Psi)|^{z=0}_{z_*}=(\Phi(z=0)+\Psi(z=0))-(\Phi(z_*)+\Psi(z_*))$.
As photons decouple from baryons and stream out of the potential wells they lose energy due to gravitational redshift so that the combination of the temperature monopole and gravitational potential, $\Theta_0+\Psi$, is representative of the effective temperature fluctuations.
The velocity of the photo-baryon fluid oscillates out of phase with radiation density and its motion along the observer line of sight causes a position dependent Doppler shift on the last scattering surface.
Combined together these two effects encode most of the information about the physical processes that leave an imprint on the CMB at recombination (see e.g.~\cite{Hu:1994uz}).
The time dependence of the gravitational potential and its change between recombination and today would lead to the ISW effect at both early and late times.
Since we are mainly interested in the ISW effect that occurs at early times we can approximate it as the difference in $\Phi+\Psi$ between recombination and today (see e.g.~\cite{Hu:1994uz}).
Fig.~\ref{Fig:RecombinationSpectrum} shows these three quantities for the $\Lambda$CDM model, a fluid DR model that is obtained by adding $f_{\rm DR}=0.6$ and a DR+AcDM model that has $f_{\rm DR}=0.6$ and $f_{\rm AcDM}=0.6$. 
In these three models all baseline parameters are left unchanged with respect to the $\Lambda$CDM ones. \nl
Adding additional radiation enhances radiation driving through the decay of the gravitational potential and hence the acoustic peaks in the CMB spectrum. 
This can be clearly seen from Fig.~\ref{Fig:RecombinationSpectrum}a where adding fluid DR enhances acoustic fluctuations in $\Theta_0+\Psi$ at last scattering, in a way that is symmetric with respect to the oscillation zero point.   The zero point is itself displaced due to baryon drag into $\Psi$ which decreases due to additional radiation. This displacement causes the familiar baryon modulation of the acoustic peaks.
The Doppler term is driven by gravitational potential decay in the same way but is not displaced by baryon drag.  The ISW contribution is a continuation of the same effect that drives the acoustic peaks but for the decay of the potential after recombination. Hence it mainly enhances fluctuations on scales larger than the first acoustic peak. The net effect on the CMB is then mainly an enhancement of power due to the driving and ISW effect from potential decay.
With respect to massless neutrinos, fluid DR, drives acoustic oscillations more efficiently due to their similarity to the tightly-coupled photons~\cite{Hu:1995en} while also slightly changing the phasing of the oscillations~\cite{Baumann:2015rya}. \nl
We now turn to the case where we add tightly coupled AcDM along with DR. If the DR has sufficient momentum ratio to the AcDM to keep the AcDM smooth relative to CDM that it will suppress the growth rate and the gravitational potential in the matter dominated era.
This suppression happens for modes that are inside the dark comoving sound horizon $r_{\rm ds}(\eta) = \int_0^\eta c_{\rm ds}(\eta') d\eta'$, where $\eta = \int dt/a$ is the conformal time, defined by the sound speed of the tightly coupled dark fluid:
\begin{equation} \label{Eq:DarkSpeedSound}
c_{\rm ds}^2 = \frac{1}{3(1+R_{\rm d})} \,,
\end{equation}
and momentum ratio:
\begin{align} \label{Eq:MomentumRatio}
R_{\rm d} =&\,\, \frac{3\rho_{\rm AcDM}}{4\rho_{\rm DR}} = \frac{3}{4(1+z)}\frac{\fACDM}{\fDR}\frac{\Omega_c h^2}{\Omega_\gamma h^2} \left[\frac{7}{8} \left(\frac{4}{11}\right)^{4/3} N_{\rm eff} \right]^{-1} \nonumber \\
   =&\,\, \frac{43883}{1+z} \left( \frac{T_{\rm CMB}}{2.725 \rm K} \right)^{-4} \left( \frac{N_{\rm eff}}{3.046} \right)^{-1} \Omega_c h^2 \frac{\fACDM}{\fDR}  \,.
\end{align}
For a given $k$ mode the transition between normal and suppressed growth occurs when:
\begin{equation} \label{Eq:DarkHorizon}
\frac{k c_{\rm ds}}{aH} = 1 \,.
\end{equation}
Note that, in the limit of a large momentum ratio, both $c_s$ and $aH$ decrease as $1/\sqrt{a}$ in matter domination so this transition depends mainly on $k$ and not $a$ so that Eq.~\ref{Eq:DarkHorizon}, evaluated today, defines the Jeans scale where fluctuations are suppressed by AcDM pressure support $k_{J} = H_0/ c_{\rm ds}(z=0)$.
After an AcDM mode enters the dark sound horizon it would start oscillating instead of growing like regular DM.
In Fig.~\ref{Fig:RecombinationSpectrum}b we compare the DR model of Fig.\ref{Fig:RecombinationSpectrum}a to the same model but with $f_{\rm AcDM}=0.6$.
As we can see this change does not introduce much change to the driving of acoustic oscillations in the effective temperature and dipole.
For $k$ modes whose potentials decay mainly before recombination the net amount of
driving is determined by their initial values which do not depend on $f_{\rm AcDM}$.  There is a 
small change in the phasing of the oscillation and a reduction in the efficiency of driving since
the impact of $\fACDM$ is no longer synchronized with the acoustic oscillations.
In addition, $\fACDM$ decreases the baryon modulation by reducing the gravitational potential
at recombination.   
It also slightly reduces the ISW effect by placing more of the potential decay before recombination.
On scales larger than the dark sound horizon at recombination, $f_{\rm AcDM}$ causes
enhanced potential decay after recombination and hence a larger ISW effect.  
Notice that while the magnitude of the effects shown in Fig.~\ref{Fig:RecombinationSpectrum} is exaggerated by the value of DR and AcDM fractions, to be readable from a figure, the scale at which the effect becomes relevant would not change if $f_{\rm DR}$ and $f_{\rm AcDM}$ scale in the same way.
We anticipate that these effects will play a key role in the cosmological constraints on the model parameters, as discussed in the next sections.

Scales that are below the dark sound horizon would also have a reduced growth rate at later times.
This decrease in growth inspires a reparametrization of the model that connects it to late time observables.
A reparametrization of the model also solves a degeneracy issue for its parameters. As $\fDR\rightarrow 0$ the dark sound horizon would go to smaller and smaller scales and AcDM would not undergo dark acoustic oscillations but rather behave as cold dark matter at all observable scales.
That is, in this limit, all parameter choices $0 \le \fACDM \le 1$ correspond to the same phenomenology as $\Lambda$CDM.
With this in mind we want to understand how the linear growth or equivalently $\sigma_8(f_{\rm DR},f_{\rm AcDM})$ behaves when all other cosmological parameters are held fixed. Notice that it is important that $H_0$ is fixed since $\sigma_8$ is the rms density fluctuation at $8h^{-1}$Mpc and $f_{\rm DR}$ at fixed $\theta_s$ will
change $H_0$.
DM at sub dark sound horizon scales would see smooth AcDM that will suppress its growth rate, in matter domination, from $a$ to $a^{1-p}$ where:
\begin{equation}
p = \frac{5 - \sqrt{1+24 (1-\fACDM) }}{4} \,.
\end{equation}
Below the Jeans scale, the growth is suppressed in matter domination as
\begin{equation}
\lim_{k\gg k_J} p_S = \left( \frac{1}{a_{\rm eq}} \right)^{-p} \,,
\end{equation}
\begin{figure}[th]
\centering
\includegraphics{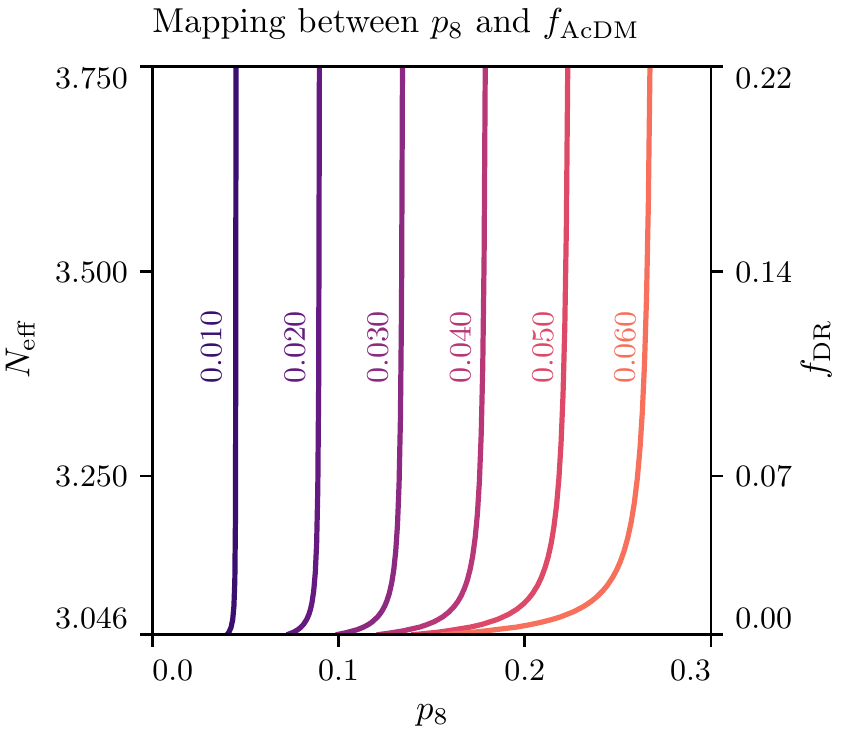}
\caption{
The level contours of the mapping between $p_8$ and $f_{\rm AcDM}$. Different colors correspond to different values of $f_{\rm AcDM}$ at fixed CDM physical density $\Omega_c h^2=0.1199$.
}
\label{Fig:P8ToACDM}
\end{figure}
and assuming that $p \ll 1$, we can expand this as:
\begin{equation}
\lim_{k\gg k_J}  p_S = -\frac{3}{5} \fACDM \ln (1/a_{\rm eq})  \,.
\end{equation}
At a fixed $k$ the growth should interpolate between the sub and super Jeans limits:
\begin{equation}
p_S \simeq -{\frac{3}{5} \fACDM \ln (1/a_{\rm eq}) }\frac{ (k/k_J)^n}{1+ (k/k_J)^n} \,,
\end{equation}
with $n>0$.
This motivates a functional form for a new parameter $p_8$ of the form:
\begin{equation} \label{Eq:DefinitionP8}
p_8 = C_1 \fACDM \frac{1}{1+[C_2 (1+R_{\rm d})^{1/2}]^n} \,,
\end{equation}
that gives the relative suppression in power as a function of $\fACDM$ and $f_{\rm DR}$.
We numerically determine the parameters of the $p_8$ parametrization by evaluating $\sigma_8(f_{\rm DR},\fACDM$ on a grid in $f_{\rm DR}$ and $\fACDM$ and we fit $p_8 \equiv \sigma_8(f_{\rm DR}, \fACDM) / \sigma_8(f_{\rm DR},0) -1$.
Note that since we do not model the $\fDR$ effects on matter-radiation equality, we define this ratio to remove the associated pure $\fDR$ effects.
For ACDM fractions smaller than $20\%$ we find that the best description of the numerical results is given by $n=2$, $C_1 = 4.5004$ and $C_2=0.002776$.
In Fig.~\ref{Fig:P8ToACDM} we show the level contours of the mapping between $p_8$ and $f_{\rm AcDM}$ at fixed CDM density.
As can be seen from Eq.~(\ref{Eq:DefinitionP8}) and Fig.~\ref{Fig:P8ToACDM} at large DR fraction $p_8$ is linearly proportional to $\fACDM$, with proportionality $C_1$.
When the DR fraction, $\fDR$, vanishes, $R_{\rm d} \rightarrow \infty$ and $p_8$ goes to zero regardless of the AcDM fraction.
This introduces a non-trivial Jacobian between the ($f_{\rm AcDM}$,$\fDR$) and ($p_8$,$\fDR$) parameter bases that has the property of singling out the $\Lambda$CDM model, mapping it to a single point in parameter space, $(0,0)$, allowing our parameter searches to focus on deviations from $\Lambda$CDM, rather than associating a large volume of parameter space to the fiducial model. \nl

\section{Data sets and methodology}  \label{Sec:DataSets}
\begin{table}
\centering
\begin{tabular}{|c| l |c|c|}
\hline
Acronym  & Data set & Year & Reference \\
\hline
 CMB  & {\it Planck} High-$\ell$ TT    & 2015 & \cite{Aghanim:2015xee} \\
 		  & {\it Planck} Low-$\ell$ TEB  & 2015 & \cite{Aghanim:2015xee} \\
 \hline
 H      & SH0ES  $H_0$                 & 2016 & \cite{Riess:2016jrr}    \\
 		  & H0LiCOW $H_0$          & 2016 & \cite{Bonvin:2016crt} \\
\hline
 WL    & CFHTLenS & 2016 & \cite{Joudaki:2016mvz}    \\
 		  & KiDS         & 2016 & \cite{Hildebrandt:2016iqg} \\
\hline
 ALL   & \begin{tabular}{@{}l@{}}  CMB +WL +H +BAO \\  JLA SN +CMB lensing \\ CMB polarization  \end{tabular} & & \cite{Aghanim:2015xee,Riess:2016jrr,Bonvin:2016crt,Joudaki:2016mvz,Hildebrandt:2016iqg,Ade:2015xua, Aghanim:2015xee,Ade:2015zua,Betoule:2014frx,Alam:2016hwk,Ross:2014qpa,Beutler:2011hx} \\
\hline
\end{tabular}
\caption{Data sets and data sets combinations used in this work.}
\label{Table:Datasets}
\end{table}
\begin{figure*}[!ht]
\centering
\includegraphics{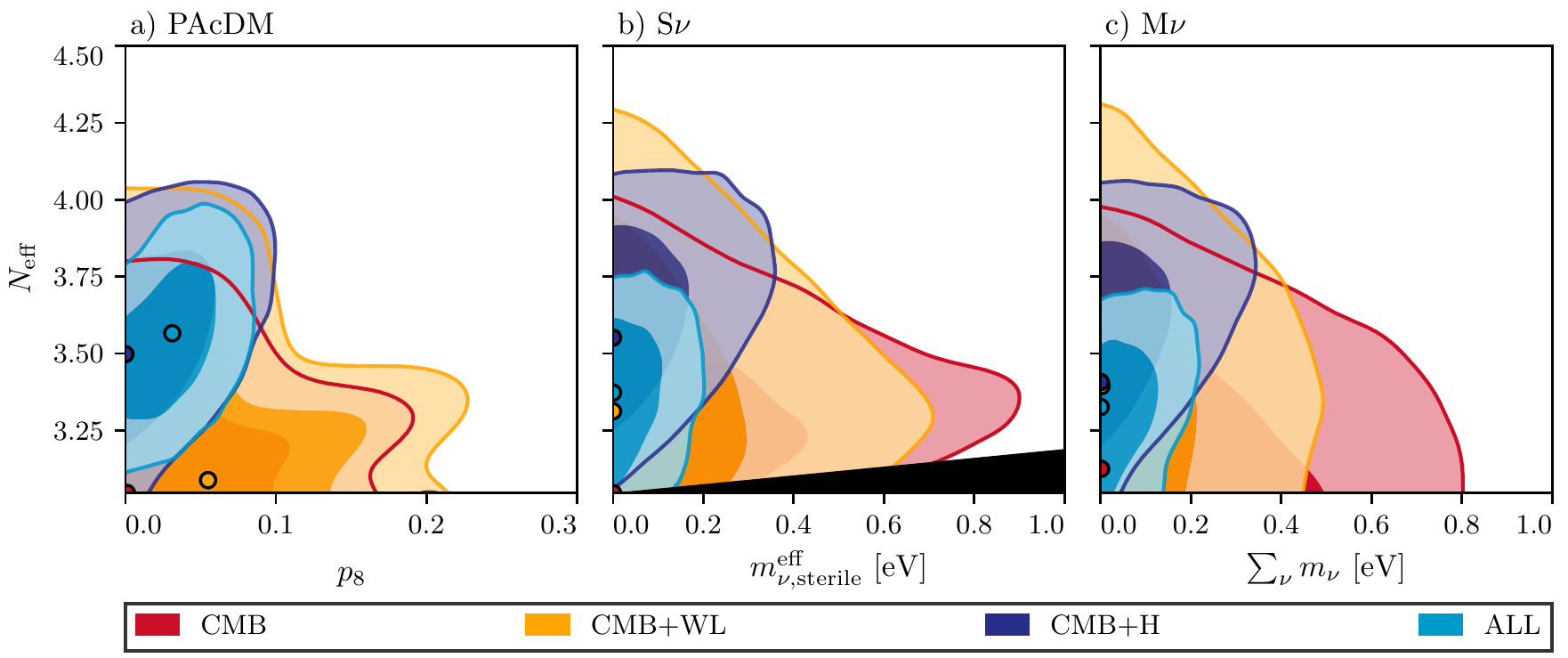}
\caption{
The marginalized joint posterior for the parameters defining the three extended models that we consider.
In all panels $N_{\rm eff}$ denotes the effective number of relativistic species, while $p_8$ indicates the growth suppression in the PAcDM case, $m_{\nu, {\rm sterile}}^{\rm eff}$ stands for the effective mass of sterile neutrinos and $\sum_{\nu} m_{\nu}$ denotes the sum of neutrino masses.
In all panels the circled points represent the best fit parameter solution for a given data set combination.
In all three panels different colors correspond to different combination of cosmological probes, as shown in legend.
The darker and lighter shades correspond respectively to the $68\%$ C.L. and the $95\%$ C.L. The black area in {\it Panel b)} shows the Dodelson-Widrow cut on the sterile neutrino effective mass.
}
\label{Fig:ExtendedParameters}
\end{figure*}

In this work we use four different combinations of data sets, as summarized in Tab.~\ref{Table:Datasets}. \nl
We use CMB measurements of {\it Planck} at both large and small angular scales.
At large angular scales we employ the {\it Planck} released joint pixel-based likelihood including both temperature and E-B mode polarization 
for the multipoles range of $\ell \leq 29$, as described in~\cite{Aghanim:2015xee}.
At smaller angular scales we use the \texttt{Plik} likelihood~\cite{Aghanim:2015xee} for CMB measurements of the TT power spectrum, as extracted from the $100$, $143$, and $217$ GHz HFI channels. 
We refer to the combination of the low-$\ell$ TEB measurements and the high-$\ell$ TT data as the CMB compilation. \nl
The data set combination including local measurements of the Hubble constant consists in: the value derived by the ``Supernovae, H0, for the Equation of State of dark energy'' (SH0ES) team~\cite{Riess:2016jrr}; measurements derived from the joint analysis of three multiply-imaged quasar systems with measured gravitational time delays, from the H0LiCOW collaboration~\cite{Bonvin:2016crt}. \nl
Our combination of weak gravitational lensing data consists in the measurements of the galaxy weak lensing shear correlation function as provided by the 
Canada-France-Hawaii Telescope Lensing Survey (CFHTLenS)~\cite{Heymans:2013fya}. 
This is a $154$ square degree multi-color survey, optimized for weak lensing analyses, that spans redshifts ranging from $z\sim0.2$ to $z\sim 1.3$.
Here we consider the reanalysis of the data as in~\cite{Joudaki:2016mvz} and we applied ultra-conservative cuts that exclude $\xi_{-}$ completely and cut the $\xi_+$ measurements at scales smaller than $\theta=17'$ for all the tomographic redshift bins. \nl
These are complemented by the tomographic weak gravitational lensing data from the $\sim450\,$deg$^2$ Kilo Degree Survey (KiDS)~\cite{deJong:2012zb,Kuijken:2015vca,Hildebrandt:2016iqg}. As in the previous case we use ultra conservative cuts to remove non-linear scales from our analysis. 
For both data sets we verified that non-linearities, as described by Halofit~\cite{Smith:2002dz}, and with the updated fitting formulas described in~\cite{Takahashi:2012em,Mead:2015yca}, where not significantly influencing the fit. \nl
For both data sets we include the modeling of intrinsic alignments, as in~\cite{Joudaki:2016mvz,Hildebrandt:2016iqg}. We do not include additional photo-$z$ uncertainties in CFHTLenS analysis as~\cite{Joudaki:2016mvz} shown that they do not qualitatively influence the results.
We do not include baryonic feedback as it should not be relevant at the scales that we are considering.
A posteriori we also observe that, by considering ultra-conservative cuts, intrinsic alignment is hardly constrained. \nl
Finally, to express the maximum tension between CMB and weak lensing probes and $H_0$ measurements we combine all these data sets together. 
We further add to the full data set combination: {\it Planck} measurements of small scales TE EE CMB polarization  power spectra~\cite{Ade:2015xua, Aghanim:2015xee}; the {\it Planck} 2015 full-sky lensing potential power spectrum~\cite{Ade:2015zua} in the multipoles range $4\leq \ell \leq 400$; the ``Joint Light-curve Analysis'' (JLA) Supernovae catalog~\cite{Betoule:2014frx}; BOSS BAO measurements in its DR12 data release~\cite{Alam:2016hwk} together with the SDSS Main Galaxy Sample~\cite{Ross:2014qpa} and the 6dFGS survey~\cite{Beutler:2011hx} BAO measurements. \nl
We use a modified version of CAMB~\cite{Lewis:1999bs} to get cosmological predictions for the PAcDM model and we sample cosmological parameters with CosmoMC~\cite{Lewis:2002ah}. We plan to make the code publicly available in the near future.
To assess whether the data sets available prefer one model over the other we perform an evidence based comparison and compute the evidence from MCMC runs with the algorithm described in~\cite{Heavens:2017afc}.  \nl
For neutrino models we consider both active and sterile variants.
When considering sterile neutrinos we define their Dodelson-Widrow (DW) mass~\cite{Dodelson:1993je} to be $m_{\rm DW} = m_{s}/\Delta N_{\rm eff}$ and $\Delta N_{\rm eff} = N_{\rm eff} -3.046$ and we bound it by imposing $m_{\rm DW} < 7 {\rm eV}$ as a prior, as in~\cite{Wyman:2013lza}, to cut the degeneracy between very massive neutrinos and CDM that is not interesting for eV scale neutrino physics.
Since we take flat priors on $m_{s}$ and $\Delta N_{\rm eff}$ and we add on top of that the DW cut we have to correct the prior volume when it is used in the evidence calculation. This volume correction factor, for the priors that we use, is ${\rm ln} \Delta V_{\rm prior} = -0.05$ that corresponds to about a $5\%$ reduction in prior volume.
When considering active neutrinos we distribute the net sum of their masses $\sum_\nu m_\nu$ equally over the active neutrinos and we do not add any prior coming from neutrino flavor oscillations. \nl
\section{Cosmological Constraints} \label{Sec:Constraints}
\begin{figure}[!ht]
\centering
\includegraphics{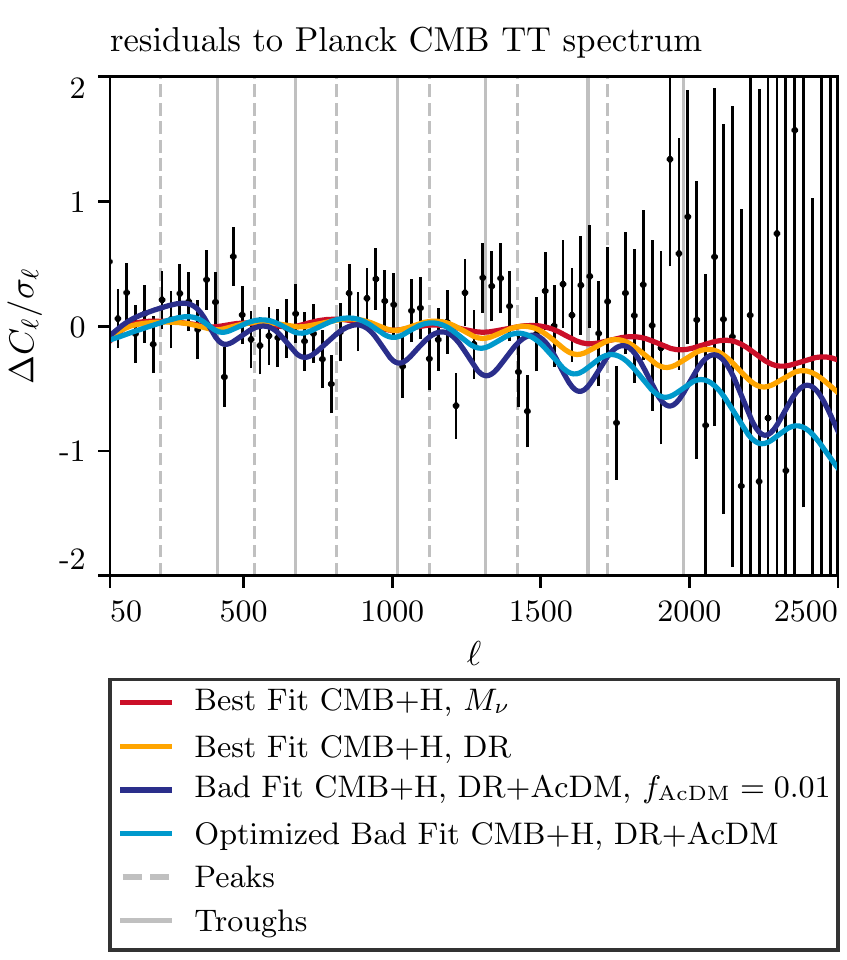}
\caption{
Residuals to {\it Planck} best fit $\Lambda$CDM temperature spectrum in units of cosmic variance per multipole. Different lines correspond to different models and parameters, as shown in legend, whose parameters are given in Tab.~\ref{Table:Parameters}.
Vertical continuous and dashed lines correspond, respectively, to the location of peaks and troughs in the best fit $\Lambda$CDM model.
Points with error bars show the residual of {\it Planck} temperature measurements.
}
\label{Fig:BestFitDecomposition}
\end{figure}
\begin{table*}
\centering
\begin{tabular}{ l | c c c c c c c c c c c c c }
\hline
Panel (a) & $\sum_{\nu}m_{\nu} ({\rm eV}) $ & $N_{\rm eff}$ & $p_8$ & $f_{\rm AcDM}$ & $f_{\rm DR} $ & $h$ & $\theta_s$ & $\Omega_bh^2$ & $\Omega_ch^2$ & $10^{-9} A_s$ & $n_s$ & $\tau$ & $S_8$ \\
\hline
BF CMB $\Lambda$CDM                  & $0.06$ & $3.046$ & $-$        & $-$        & $-$        & $0.672$ & $1.04106$ & $0.02219$ & $0.1199$ & $2.204$ & $0.965$ & $0.079$ & $0.467$ \\
BF CMB+H, $M_{\nu}$                     & $0.00$ & $3.496$ & $-$        & $-$        & $-$        & $0.722$ & $1.04051$ & $0.02277$ & $0.1231$ & $2.316$ & $0.989$ & $0.100$ & $0.459$ \\
BF CMB+H, DR                                 & $0.06$ & $3.494$ & $0.0$    & $0.0$     & $0.14$ & $0.720$ & $1.04178$ & $0.02293$ & $0.1237$ & $2.263$ & $0.977$ & $0.102$ & $0.450$ \\
 \begin{tabular}{@{}l@{}}  Bad Fit CMB+H, \\  \hspace{0.2cm} DR + AcDM  \end{tabular} & $0.06$ & $3.494$ & $0.055$  & $0.012$ & $0.14$ & $0.720$ & $1.04178$ & $0.02293$ & $0.1237$ & $2.263$ & $0.977$ & $0.102$ & $0.424$ \\
 \begin{tabular}{@{}l@{}}  Optimized Bad Fit  \\  \hspace{0.2cm} CMB+H, DR + AcDM  \end{tabular} & $0.06$ & $3.494$ & $0.055$  & $0.012$ & $0.14$ & $0.713$ & $1.04216$ & $0.02267$ & $0.1257$ & $2.311$ & $0.982$ & $0.108$ & $0.441$ \\
\hline
\end{tabular}
\caption{Parameters for the models in Fig.~\ref{Fig:BestFitDecomposition}.
} 
\label{Table:Parameters}
\end{table*}

We confront the PAcDM and the active/sterile neutrino models with the data sets described in the previous section and in Fig.~\ref{Fig:ExtendedParameters} we show the marginalized posterior confidence contours of the parameters defining the three extended models. \nl
In particular, in Fig.~\ref{Fig:ExtendedParameters}a, we show the results for PAcDM. As we can see the CMB posterior has two main branches: one that corresponds to parameter combinations resulting in high $N_{\rm eff}$ or DR content, with a small fraction of AcDM and correspondingly a small suppression of late time growth; a second branch that corresponds to a lower DR content but significantly more AcDM, resulting in a stronger growth suppression.
Noticeably the solution corresponding to both a high DR/AcDM content, giving a higher Hubble constant and stronger growth suppression, is disfavored by CMB measurements. The two branches require incompatible changes to the cold dark matter and baryon densities to compensate the DR and AcDM changes to the acoustic peaks. \nl
To further interpret this exclusion we analyze the parameter dependence of the best fit models
given in Tab.~\ref{Table:Parameters}.   In Fig.~\ref{Fig:BestFitDecomposition} we compare the model predictions to the {\it Planck} data points in terms of their deviation with respect to the {\it Planck} TT best fit $\Lambda$CDM model
and in units of its cosmic variance per multipole $\sigma_{\ell}= \sqrt{2/(2 \ell+1)} C_{\ell}^{{\rm bf}(\Lambda{\rm CDM})}$.  
Since active and sterile neutrino cases behave similarly, we study the active case.  \nl
In Fig.~\ref{Fig:BestFitDecomposition}, we show the best fit solutions to CMB and H measurements in the massive neutrino and PAcDM cases. 
Among the two models the neutrino one provides a slightly better fit to the CMB spectrum with respect to the PAcDM case.
Moreover, in the PAcDM case the minimization of the chi square prefers $f_{\rm AcDM}=0$
and so we refer to this case as the best fit DR model. Similarly in the neutrino case, the best fit solution prefers $\sum m_\nu=0$.
In addition to these models we show two $\fACDM$ solutions: the first is the ``bad fit" obtained by taking the PAcDM best fit CMB+H parameters and raising the AcDM fraction to $f_{\rm AcDM}=0.01$, the best fit value preferred by CMB+WL measurements, without re-optimizing other parameters; the second is the ``optimized bad fit" obtained by fixing $f_{\rm DR}$ and $f_{\rm AcDM}$ to the values of the previous case and re-optimizing all other parameters. \nl
\begin{figure*}[!t]
\centering
\includegraphics{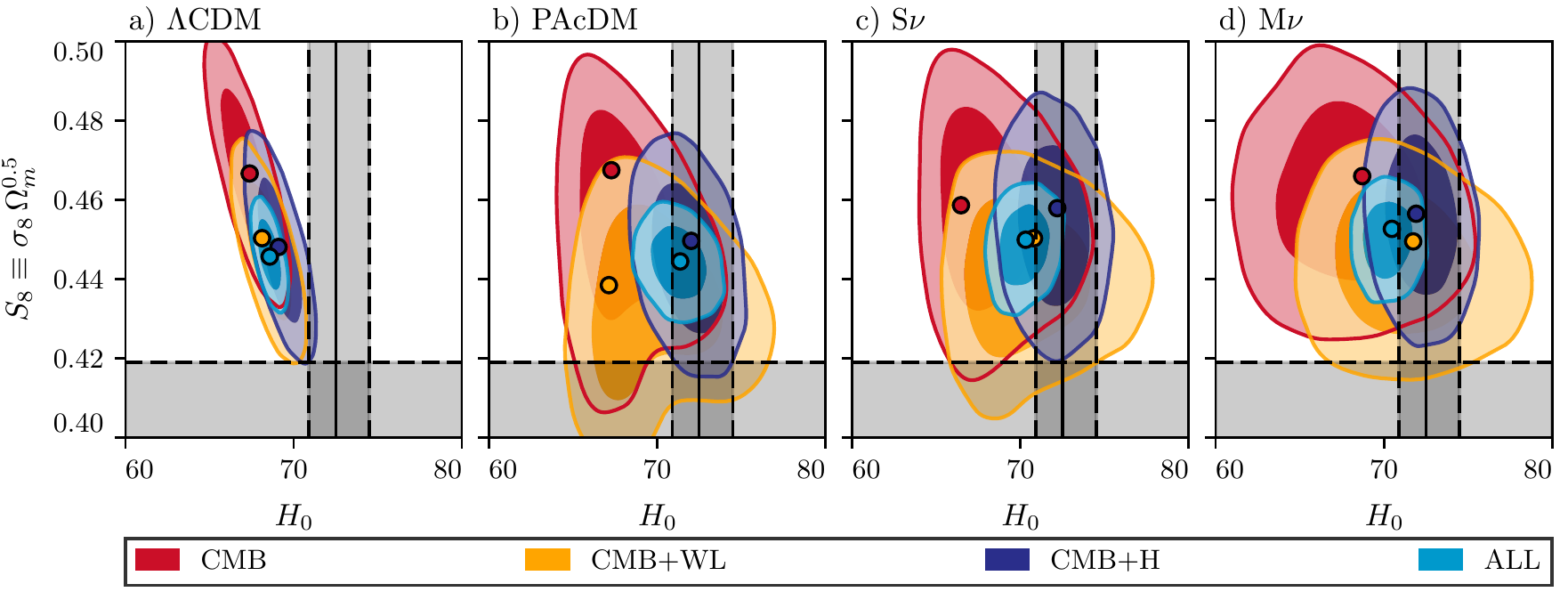}
\caption{The marginalized joint posterior for the weak lensing inferred amplitude of scalar perturbations $S_8\equiv \sigma_8\Omega_m^{0.5}$ and the present day value of the Hubble constant $H_0$  {in units of km/s/Mpc} for the different models that we consider.
In all panels the circled points represent the best fit parameter solution for a given data set combination.
In all three panels different colors correspond to different combination of cosmological probes, as shown in legend. 
The darker and lighter shades correspond respectively to the $68\%$ C.L. and the $95\%$ C.L.. 
The continuous black line shows the best fit $H_0$ and $S_8$ values, as obtained by fitting respectively the H and WL data set alone, within the $\Lambda$CDM model. The black dashed lines and shaded area indicate the $68\%$ C.L. region.
}
\label{Fig:DatasetResults}
\end{figure*}
As we can see in  Fig.~\ref{Fig:BestFitDecomposition}, the best fit solution in both the DR and neutrino cases shows nearly no residuals for well measured scales $\ell < 1500$.
This is achieved by counteracting the effects of the enhanced potential decay on radiation driving and baryon drag by raising the CDM and baryon density 
and counteracting those of damping by tilting the primordial spectrum (see Tab.~\ref{Table:Parameters}).   Since DR results in more radiation driving than neutrinos, all parameters are readjusted to get nearly the same residuals at $\ell < 1500$ while 
slightly penalizing higher multipoles where {\it Planck} errors increase. 
Thus DR is slightly less efficient at raising $H_0$ than neutrinos. \nl
When we raise the AcDM fraction from the best fit DR solution, prominent oscillatory residuals relative to $\Lambda$CDM appear corresponding to a shift in  the phasing and sharpening of the acoustic peaks. The former appears as the out of phase component with respect to the peaks and troughs and the latter as in phase component. The smoother changes corresponding to an overall enhancement of power around the first peak and a suppression
of power beyond it can be mainly compensated by tilt in the spectrum.   
Even though the background remains unchanged, increasing the AcDM fraction changes the
driving of acoustic oscillations and hence their phase relative the photon-baryon sound horizon.
The corresponding decay of the potential also reduces gravitational lensing thereby reducing
its smoothing effect on the peaks. \nl
Removing the residuals to improve the fit requires raising $\Omega_c h^2$ which makes the 
acoustic peaks smoother by decreasing radiation driving and increasing gravitational lensing 
\cite{Aghanim:2016sns,Obied:2017tpd}
as well as changing $\theta_s$ and $\Omega_b h^2$ to match the phasing (see Tab.~\ref{Table:Parameters}, optimized bad fit).  Even so, residuals increase through the damping tail, beyond the region well-measured by {\it Planck}.   Experiments targeted at measuring CMB temperature fluctuations to cosmic variance limit at small angular scales can potentially distinguish between these two scenarios.
More importantly for tensions, increasing $\Omega_c h^2$ raises
$S_8$ to nearly its $\Lambda$CDM value and slightly lowers $H_0$  (see Tab.~\ref{Table:Parameters}). 
      It is this problem that prevents PAcDM from simultaneously raising
$H_0$ and lowering $S_8$ and causes the two different branches of solutions. \nl
\begin{figure*}[!t]
\centering
\includegraphics{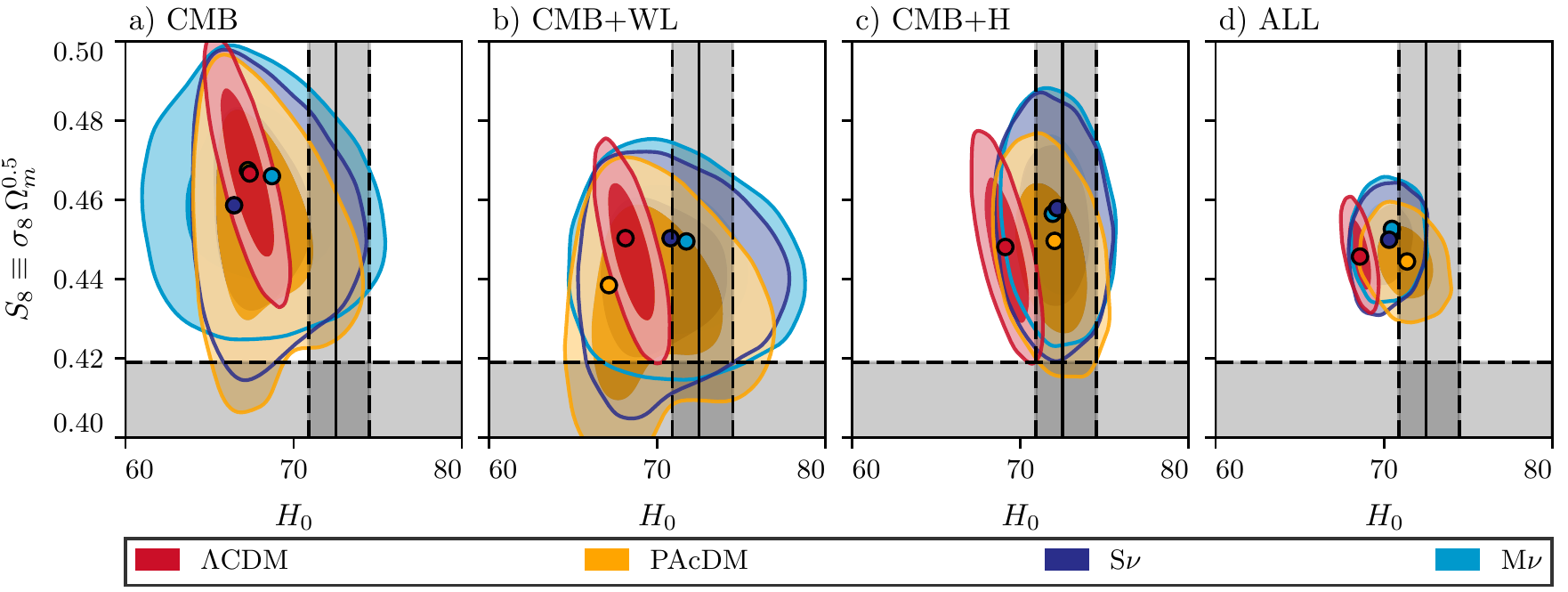}
\caption{The marginalized joint posterior for the weak lensing inferred amplitude of scalar perturbations $S_8\equiv \sigma_8\Omega_m^{0.5}$ and the present day value of the Hubble constant $H_0$ {in units of km/s/Mpc} for the different data set combinations that we consider.
In all panels the circled points represent the best fit parameter solution for a given data set combination.
In all three panels different colors correspond to different models, as shown in legend. 
The darker and lighter shades correspond respectively to the $68\%$ C.L. and the $95\%$ C.L.. 
The continuous black line shows the best fit $H_0$ and $S_8$ values, as obtained by fitting respectively the H and WL data set alone, within the $\Lambda$CDM model. The black dashed lines and shaded area indicate the $68\%$ C.L. region.}
\label{Fig:ModelResults}
\end{figure*}
The same multi-modal distribution that shows clearly in the PAcDM case is present, to some extent, in the neutrino cases but is not as manifest at the posterior level and only shows when adding Hubble constant measurements.
In the PAcDM model the branch with a high AcDM content is favored by WL observables, as the posterior extends in that direction as this data set is added to the CMB one.  The same is not true in the active and sterile neutrino cases, where the posterior is not favoring larger neutrino masses with WL data.
Neutrino mass is less efficient in suppressing structure since neutrinos are still relativistic at high redshift but this advantage of PAcDM is not manifest on the high $H_0$ branch. \nl
When combining the two data sets the one with the strongest tension statistical significance drives the posterior toward its branch. In this case we can see that the Hubble constant tension is dominating the ALL data set combination leaving $S_8$ nearly unchanged from $\Lambda$CDM. \nl
The ALL data set combination contains other data sets for completeness. We checked all different data combinations separately and their addition does not qualitatively influence the physical picture. The uncertainties on all parameters are just a factor tighter while the best fit results are largely unaffected and we do not observe significant changes in degeneracy directions. \nl
This analysis is further confirmed by Fig.~\ref{Fig:DatasetResults} where we show the performances of the four models on the $S_8$ and $H_0$ ``tension'' plane.
The main feature that the models should be fitting is the difference between $\Lambda$CDM {\it Planck} posterior and the area where the $H_0$ and $S_8$ measurements cross, that we shall refer to as the ``target'' parameters.
The $\Lambda$CDM model is constrained to move along the degeneracy where changes in the sound horizon and distance to recombination compensate to leave angular scales fixed, $\Omega_m h^3\approx$\,const.\ so that the effect of both H and WL measurements pushes the data posterior in the same direction and the joint posterior is a compromise between all the results.
The PAcDM model instead clearly shows the two branches, one going in the direction of easing the tension with weak lensing measurements, the other going in the direction of the Hubble tension while moving along the diagonal toward the target is disfavored.
This same effect is present but weaker in the sterile neutrino case and even weaker in
the active case. \nl
Fig.~\ref{Fig:ModelResults} shows the complementary view of comparing how well
the different models fit the same data combinations and we can use this as a guidance to understand the model selection results in Tab.~\ref{Table:ModelSelection}.
When we consider CMB data alone the extended models enlarge the error bars on $S_8$ and $H_0$ while not moving significantly the best fit. This is reflected by the evidence ratio disfavoring the three models while the best fit chi square shows small or no improvement. 
Notice that the three models have different priors so that the Occam's razor factor in the evidence ratio is penalizing the three extensions differently. \nl
When adding WL, in Fig.~\ref{Fig:ModelResults}b, the all models move down in $S_8$ with PAcDM allowing the lowest values and having a best fit on the branch with high $\fACDM$ and low $H_0$.  Conversely, neutrino models favor relatively higher $H_0$ and though lower
$S_8$ are allowed, the best fit remains at nearly the $\Lambda$CDM  level.
\nl
When adding the H data set on top of the CMB one,  both PAcDM and neutrinos move 
towards higher $H_0$ but now neither lower $S_8$ much compared with $\Lambda$CDM.   The neutrino models are also slightly more efficient than PAcDM in fitting $H_0$ measurements.
As opposed to fluid radiation, free-streaming radiation sources significant anisotropic stress that induces a phase shift in CMB anisotropies~\cite{Baumann:2015rya} thus allowing for slightly larger values of $H_0$.
The model selection results reflect this and the two neutrino models are the only one favored, with respect to the $\Lambda$CDM model, by the evidence comparison and Hubble constant measurements. \nl
When combining ALL datasets, PAcDM and neutrino models favor very similar
values of $H_0$ and $S_8$ with the latter nearly the same as in $\Lambda$CDM.
PAcDM allows  slightly lower values of $S_8$ while keeping $H_0$ slightly closer to local measurements.
In the evidence based comparison PAcDM is correspondingly performing slightly better than the other two models reaching a draw with $\Lambda$CDM. \nl
On the other hand, since $S_8$ is only marginally lowered from its $\Lambda$CDM value the tension between CMB, Hubble constant and Weak Lensing measurements is not simultaneously relieved in the three extended models.
The improvement in best fit, shown in Tab.~\ref{Table:ModelSelection}, for a two parameter models, correspond to roughly a two sigma feature and is driven by Hubble measurements as this is the confidence level at which $N_{\rm eff}$ deviates from its standard value. \nl
While these results are influenced by the relative statistical significance of the two tensions, that selects the posterior branch that the model will occupy and the one that the model will sacrifice, a generic feature is that these models cannot relieve both tensions at the same time without sacrificing the goodness of fit to CMB measurements.

\begin{table}
\centering
\begin{tabular}{|c| l |c|c|c|c|}
\hline
Combination & Model & $\chi^2_{\rm max}$ & ${\rm ln} \mathcal{E}$ & $\Delta \chi^{2}_{\Lambda{\rm CDM}}$ & $\Delta {\rm ln} \mathcal{E}_{\Lambda{\rm CDM}}$\\
\hline
 CMB  & $\Lambda$CDM & $11263$ & $-5710.5$ & $-$ & $-$      \\
 		  & PAcDM              & $11262$ & $-5714.1$ & $-1$ & $+3.6$ \\
 		  & S$\nu$                   & $11263$ & $-5711.7$ & $0$ & $+1.2$ \\
 		  & M$\nu$                   & $11262$ & $-5712.2$ & $-1$ & $+1.7$ \\
\hline
 CMB + WL    & $\Lambda$CDM & $11414$ & $-5794.6$  & $-$ & $-$     \\
 		  			 & PAcDM              & $11412$ & $-5797.2$  & $-2$ & $+2.6$ \\
 		  		     & S$\nu$                   & $11412$ & $-5795.4$  & $-2$ & $+0.8$ \\
 		  		     & M$\nu$                   & $11413$ & $-5796.0$ & $-1$ & $+1.4$ \\
 \hline
 CMB + H    & $\Lambda$CDM & $11314$ &  $-5736.4$ & $-$ & $-$     \\
 		  			 & PAcDM              & $11308$ & $ -5737.1$ & $-6$ & $+0.7$ \\
 		  			 & S$\nu$                   & $11306$ & $-5734.5$  & $-8$ & $-1.9$ \\
 		  			 & M$\nu$                   & $11305$ & $-5734.4$ & $-9$ & $-2.0$ \\
\hline
 ALL   & $\Lambda$CDM & $14034$ & $-7170.1$ & $-$ & $-$     \\
 		  & PAcDM              & $14028$ & $-7170.1$ & $-6$ & $+0.0$ \\ 
 		  & S$\nu$                   & $14030$ & $-7171.0$ & $-4$ & $+0.9$ \\ 
 		  & M$\nu$                  & $14030$ & $-7170.8$ & $-4$ & $+0.7$ \\  
\hline
\end{tabular}
\caption{Model selection results for different models and different data set combinations. Notice that positive values of $\Delta {\rm ln} \mathcal{E}_{\Lambda{\rm CDM}}$ and $\Delta \chi^{2}_{\Lambda{\rm CDM}}$ are in favor of the $\Lambda$CDM model.}
\label{Table:ModelSelection}
\end{table}
\section{Outlook} \label{Sec:Conclusions}
We have investigated the cosmology and observational implications of a tight coupling between part of the dark matter sector and a dark radiation sector.
This model of Partially Acoustic Cold Dark Matter (PAcDM) was proposed in~\cite{Chacko:2016kgg} as a possible way to solve the tension between CMB observations, local measurements of the Hubble constant and Weak Lensing data.  
We have shown however that the PAcDM model induces non-trivial effects on the cosmic microwave background changing the behavior of radiation driving of acoustic oscillations, the early ISW effect, baryon modulation and gravitational lensing which compromise its ability to resolve both
tensions simultaneously. \nl
We tested the model with state of the art cosmological data sets including measurements of the Hubble constant from strong lensing time delays, weak lensing from the CFHTLenS and KiDS surveys. To the best of our knowledge this is the first cosmological application of these latter two sets together. \nl
Based on our analysis of the CMB behavior we have shown that the PAcDM model has two main posterior branches: one in which the Hubble constant is raised by increasing the fraction of dark radiation with a marginal suppression in the growth of structures; a second branch, with a significant fraction of AcDM, that correspond to a strong decrease in the growth at the expense of a constant value for the Hubble constant.
While the former parameter branch can relieve the tension with Hubble constant measurements, the latter would reduce the significance of the discrepancy with weak lensing surveys.
We found, however, that the combination of the two parameters that is needed to relieve
both tensions, as well as the benchmark parameters proposed in~\cite{Chacko:2016kgg}, is disfavored by CMB observations. \nl
More specifically, while increasing the AcDM fraction at a high value of dark radiation that
relieves the Hubble constant tension on its own does lower $S_8$, it
also changes the phasing and sharpness of the acoustic peaks.   These require the cold
dark matter density to be raised to compensate which then drives $S_8$ back up.  While
still allowed, there is no evidence for PAcDM deviations from $\Lambda$CDM when both
tensions are considered. \nl
Neutrino models behave similarly but with somewhat different adjustments of parameters.
These adjustment, while producing near degeneracies for the {\it Planck} data, become clearly
distinguishable on smaller scales.
Measurements from small scales CMB experiments like SPTPol~\cite{Henning:2017nuy}, ACTPol~\cite{Louis:2016ahn} or CMB-S4~\cite{Abazajian:2016yjj} could potentially definitively tell PAcDM models  apart from $\Lambda$CDM and neutrino alternatives with precise measurements of the CMB damping tail.
Complementing this analysis with measurements from the Dark Energy Survey~\cite{Abbott:2017wau} could potentially tell whether the marginal suppression in growth found in models relieving the tension with Hubble measurements could be enough to fit also weak lensing observations.

\acknowledgments
We thank Chen He Heinrich, Meng-Xiang Lin, Andrew J. Long, Pavel Motloch and Samuel Passaglia for useful comments.
MR is supported by U.S. Dept. of Energy contract DE-FG02-13ER41958 and LTW
by DE-SC0013642.
WH was additionally supported by the Kavli Institute for Cosmological Physics at the University of Chicago through grants NSF PHY-0114422 and NSF PHY-0551142, NASA ATP NNX15AK22G and the Simons Foundation. This work was completed in part with
resources provided by the University of Chicago Research Computing Center.

{\it Note added.--} Recently, a related paper~\cite{Buen-Abad:2017gxg} commented upon similar ideas. 
Our work focuses on a different combination of datasets, while also being more conservative about the inclusion of non-linear scales.
Notably we do not include cluster abundance measurements.
Presenting different data combinations leads to a different perspective on the necessity and the effectiveness of relieving tension between the CMB, Hubble constant and local structure.

\bibliography{biblio}
\end{document}